\documentclass[1p]{elsarticle}
\usepackage{lipsum}
\makeatletter
\def\ps@pprintTitle{%
 \let\@oddhead\@empty
 \let\@evenhead\@empty
 \def\@oddfoot{}%
 \let\@evenfoot\@oddfoot}
\makeatother
\usepackage{moreverb}
\usepackage{lineno,hyperref,amsmath,dsfont,graphicx,latexsym,color,multirow,amssymb}
\usepackage{graphicx}
\usepackage{subcaption}
\usepackage{pgfplots}
 \modulolinenumbers[5]
 
\newcommand{\x}{{\bf x}}

\journal{Nonlinear Analysis: Hybrid Systems}

\begin{document}

%\runningheads{A. Ba\~nos and  M. A. Dav\'o}{Stability of time-delay reset systems}

%\noreceived{}
%\norevised
%\noaccepted
\begin{frontmatter}

\title{Stability of time-delay reset systems with a nonlinear and time-varying base system\tnoteref{t1,t2}}
\tnotetext[t2]{\copyright {\ }2016. This manuscript version is made available under the CC-BY-NC-ND 4.0 license http://creativecommons.org/licenses/by-nc-nd/4.0/}
\author{A. Ba\~nos\corref{mycorrespondingauthor}}
\cortext[mycorrespondingauthor]{Corresponding author}
\ead{abanos@um.es}
\author{M. A. Dav\'o}
\address{Dpt. Computer and Systems Engineering, University of Murcia, 30100 Murcia, Spain}

\begin{abstract}
This work is devoted to investigate the stability properties of time-delay reset systems. We present a Lyapunov-Krasovskii proposition, which generalizes the available results in the literature, providing results for verifying the stability of time-delay reset systems with nonlinear and time-varying base system. We demonstrate the applicability of the proposed results in the analysis of time-delay reset control systems, and an illustrative example with nonlinear, time-varying base system.
\end{abstract}


\end{frontmatter}

\section{Introduction}

The investigation of reset systems was started more than fifty years ago with the seminal work of Clegg \cite{clegg}, and \textcolor{black}{carried on with} a series of works by Horowitz and coworkers (\cite{KriHor, HorRos}). The main motivation for the study of reset systems arises from reset control (\cite{BanBarBook}), since reset compensation may achieve fast and robust control solutions for problems under linear limitations, which as it is well-known are particularly severe in the case of control systems with time-delays (\cite{astrom}). A large number of works have shown the advantages of reset control over linear control (\cite{beker2001,davo,moreno, BanBarBook}). 

\textcolor{black}{The term {\em reset system} was first introduced by Hollot, Chait, and coworkers (\cite{BekHoll}) to denote a ``linear and time invariant system with (state-dependent) mechanisms and laws to reset their states to zero". Two distinctive characteristics of reset systems are that the resetting law is state-dependent, and that (some) states are reset to zero. Therefore, reset systems can be considered as a special type of impulsive/hybrid systems, in which the system state (or a part of it) is instantaneously zeroed out at those instants in which the system solution intersects some reset set. On the other hand, time-delay systems (see the monographs \cite{hale,gu,F2014}, or for example the works \cite{fridman,Cao2004,Cao2006,GP2006,SG2013}) are a natural target for reset control. As a result, time-delay reset systems have a clear interest in control practice and have been an active research topic in the last decade (\cite{BanBar,BarBan, BanBarBook}).  \\
In addition, it should be noted that impulsive systems (without time-delays) have been a very active research topic in different areas of mathematics and systems theory (\cite{BS1989,LBS1989,HCN2006,MHD2008}), where the research effort has been concentrated in systems with impulses at fixed/variable instants. However, as it has been discussed in (\cite{BanBarBook, r_BM2012}) reset systems (with a time invariant base system) are a special case of autonomous impulsive systems, a much less developed research topic in the literature. On the other hand, time-delay impulsive systems, or more specifically impulsive functional differential equations, have been investigated in a number of works.  Again, most of the research effort has been concentrated in systems with impulses at fixed instants (\cite{BL2000,delasen,impulsivo2004,impulsivo2007-2,impulsivo2008,SMZ2005}), and to a lesser extent in systems with impulses at variable times (\cite{impulsivo2007-3}). Note that the vast majority of stability results for impulsive functional differential equations uses Lyapunov-Razumikhin techniques (see \cite{S2009} and references therein).}
%IVANKA: Impulsive functional differential equations are a natural generalization of impulsive ordinary differential equations (without delay) and of functional differential equations (without impulses). At the present time the qualitative theory of such equations un- dergoes rapid development. Many results on the stability and boundedness of their solutions are obtained. It is natural to ask whether we can find a systematic account of recent developments in the stability and boundedness theory for impulsive func- tional differential equations.
%
%in the past decades, the importance of impulsive systems has been highlighted and a number of results on impulsive functional differential equations haeve appeared. In spite of the fact that significant progress on the stability of impulsive dynamical systems with time-delay has been made in the last decade (\cite{impulsivo2007,impulsivo2009,impulsivo2011,impulsivo2011-2,impulsivo2012,impulsivo2014})%(\cite{impulsivo2004,impulsivo2006,impulsivo2007,impulsivo2007-2,impulsivo2007-3,impulsivo2008,impulsivo2009,impulsivo2010,impulsivo2011,impulsivo2011-2,impulsivo2012,simpulsivo2014}), 
%, most of the results are focused on systems with impulses at fixed/variable instants; and thus,  leaving reset systems, modeled as autonomous impulsive systems, almost unexplored. 

%The objective of
This work is focused on the internal stability analysis of time-delay reset systems by using a Lyapunov-Krasovskii approach, with the goal of obtaining conditions formulated by using Lyapunov-Krasovskii functionals. \textcolor{black}{To the knowledge of authors, all the previous published results from the impulsive functional differential equations literature are restricted to systems with impulses at fixed times, and thus they are not applicable to our case, or are based on Lyapunov-Razumikhin techniques. In spite of that,} the Lyapunov-Krasovskii approach has been already investigated in the area of reset systems; 
in particular, a delay-independent condition is obtained in \cite{BanBar} for reset control systems, and, in addition, extension to delay-dependent conditions is given in \cite{BarBan}, and more recently  in \cite{Pri}, considering a more general resetting law (anticipative reset). Also,  quadratic stability of time-delay reset control systems with uncertainty in the resetting law has been analyzed in \cite{Guo}. In general, a previous work \cite{DavBan} suggests that it is necessary a deeper analysis of Lyapunov-Krasovskii functionals, focusing into the necessity of obtaining less restrictive conditions.  A first attempt to obtain less restrictive conditions is \cite{DavBan}, where it has been proposed criteria based on bounded increments of the functional after the reset instants. In addition, input-output stability has been investigated in \cite{MerDav}, based on the previous Lyapunov-Krasovskii results, and in \cite{MerCar}, based on passivity properties of reset systems \cite{CarBanArj}, and the IQC framework. 

To the authors knowledge, all the previous published results are about reset systems with a linear and time-invariant (LTI) base system, and most of them are based on the existence of a Lyapunov-Krasovskii theorem; and therefore, the proof of that theorem has been only sketched, making the generalization to nonlinear and time-varying systems challenging.  
%%%%%%%%%%%%
\textcolor{black}{On the other hand, some preliminary works \cite{teel,teel2, r_BPTZ2014} have extended the hybrid inclusion model developed in \cite{c13} to investigate hybrid systems with time-delays, based on a generalized concept of solutions.  
In addition, the results in \cite{teel2} provide sufficient conditions for the stability analysis of hybrid systems with time-delays, using Lyapunov-Razumikhin functions, but application to reset control system is still unexplored. On the other hand, although \cite{r_BPTZ2014} approaches the case of reset systems, it deals with a restrictive class of reset systems in which the time-delay only affects a part of the state and thus its applicability is very limited in practice.}

The main result of this work is the development of a Lyapunov-Krasovskii theorem for reset systems, \textcolor{black}{in which the 
resetting law is generalized in the sense that the state may be reset to a non-zero value after a reset action, and in addition the base system is nonlinear and time-varying.} In this way, this work is devoted to provide a formal and complete proof, and then to analyze reset systems with a LTI base system as a particular case. 
% of a generalized theorem for time-delay reset systems with nonlinear and time varying base system.
The paper is structured as follows. After formally stating the problem in Section \ref{section:preliminares}, the main stability result is given in Section \ref{section:mainresult}. In Section \ref{section:application}, two application cases of the stability result are shown; firstly, a general reset system with a LTI base system and a single time-delay; and secondly, a particular reset system with a nonlinear and time-varying base system. The work concluded in Section \ref{section:conclusion} with some final remarks.\\

{\em Notation:} $\mathds{R}$ is the set of real numbers, $\mathds{R}^+$ is the set of non-negative real numbers, $\mathds{R}^n$ is the n-dimensional euclidean space, where $\|\mathbf{x}\|$ is the euclidean norm for $\mathbf{x} \in \mathds{R}^n$, and $(\x, {\bf y })$, with column vectors $\x \in \mathds{R}^n$ and ${\bf y} \in \mathds{R}^m$, denotes the column vector $ \left( \ \x^\top \ \ {\bf y}^\top  \ \right)^\top$. ${\cal PC}([a,b],F)$ is the set of piecewise continuous functions from $[a,b]$ to $F$, that is the set of functions that are continuous on $[a,b]$ except in a finite number of points $(t_k)_{k=1}^N$, and with a norm $\|\mathbf{\psi}\| = \sup_{\theta \in [a,b]} \| \psi(\theta)\|$. $\mathcal{R}(A)$ and $\mathcal{N}(A)$, for a matrix $A\in \mathds{R}^{n\times m}$, stand for the column space and the null space of A, respectively. $diag(A,B)$ is a block diagonal matrix composed by the matrices $A\in \mathds{R}^{n\times n}$ and $B\in \mathds{R}^{m\times m}$. For a symmetric matrix $A \in \mathds{R}^{n\times n}$, $\lambda_m(A)$ and $\lambda_M(A)$ stand for the minimum and maximum eigenvalue, respectively.

\section{Preliminaries}
\label{section:preliminares}
Consider a state-dependent time-delay reset system given by the impulsive
differential equation
\begin{equation}
\label{eq:resetsystem}
  \left\{ \begin{array}{llll}
            \dot{\mathbf{x}}(t) =  \mathbf{f}(t,\mathbf{x}_t),  &   \mathbf{x}(t) \notin {\cal M},   \\
             {\mathbf{x}}(t^+) = \mathbf{I}(t,\mathbf{x}_t), &  \mathbf{x}(t) \in {\cal M},\\
             \mathbf{x}(t) =  \phi(t-t_0), & t \in [t_0-h,t_0],
\end{array}\right.
\end{equation}
\noindent where $t_0\in \mathds{R}^+$ is the initial instant, $\mathbf{x}(t) \in \mathds{R}^n$ is the system state at the instant $t \in [t_0-h,\infty)$, $\mathbf{x}_t \in {\cal PC}([-h,0],\mathds{R}^n)$ is the distributed state at the instant $t \in [t_0,\infty)$, that is $\mathbf{x}_t(\theta) = \mathbf{x}(t+\theta)$ for $\theta \in [-h,0]$, and the initial condition is a function $\phi \in {\cal PC} ([-h,0],\mathds{R}^n)$. Henceforth, it will be denoted ${\cal PC} ={\cal PC} ([-h,0],\mathds{R}^n)$ for notational simplicity. 
%\textcolor{black}{It will be assumed that both $\mathbf{f}:\mathds{R}^+ \times {\cal PC} \rightarrow \mathds{R}^n$ and $\mathbf{I}:\mathds{R}^+ \times {\cal PC} \rightarrow \mathds{R}^n$ map $\mathds{R}^+ \times (\text{bounded sets of }{\cal PC} )$ into bounded sets of  $\mathds{R}^n$.} 
It is considered that the reset is applied at instants $t_k \in [t_0,\infty)$ if $\mathbf{x}(t_k) \in {\cal M}$, \textcolor{black}{where ${\cal M} \subset \mathds{R}^n$ is the reset set. %, and it is defined by
%$\mathbf{f}: \mathds{R}^+ \times {\cal PC} \rightarrow \mathds{R}^n$ is Lipshitz continuous in $\psi$, that is there exists a real constant $K\geq 0$ such that $\|\mathbf{f}(t,\psi_1) - \mathbf{f}(t,\psi_2) \| \leq K \|\psi_1 - \psi_2\|$ for any $\psi_1, \psi_2 \in \cal{PC}$ and any $t \in [t_0, \infty)$, the reset function $\mathbf{I}:\mathds{R}^+ \times {\cal PC} \rightarrow \mathds{R}^n$ is assumed to be continuous, 
%\textcolor{red}{
%\begin{equation}\label{eq:resetset}
%{\cal M} \ = \ \left\{  \mathbf{x} \in \mathds{R}^n : \ g(\x) = 0 \right\}
%\end{equation}
%for some function $g: \mathds{R}^{n} \rightarrow \mathds{R}$.} 
It is assumed that reset instants are {\em well-posed}, that is} for any initial condition $\phi \in {\cal PC}$ there exists a finite or infinite sequence of well defined reset instants $\mathds{T}_\phi = (t_1, t_2, \cdots)$, such that they are distinct and satisfy  $t_0<t_1 < t_2 < \cdots$; \textcolor{black}{and  also that reset instants are Zeno-free,} that is if the reset instants sequence is infinite then $t_N \rightarrow \infty$ as $N \rightarrow \infty$. Otherwise, as well as in the case of free-delay impulsive dynamical systems (\cite{HCN2006}), pathological behaviors like beating and existence of Zeno solutions \textcolor{black}{may be present. A simple manner to guaranty that reset instants are well-posed and Zeno-free is to use time regularization} (see for example \cite{BanBarBook}), which means that reset instants satisfy $t_{k+1}-t_k> \Delta$, $k=0,1,2,\ldots$, \textcolor{black}{for some $\Delta >0$} (the initial instant $t_0$ is not a reset instant). 

Thus, the existence and uniqueness of solutions of the reset system (\ref{eq:resetsystem}) follows from the existence and uniqueness of the following initial-value problem
\begin{equation}\label{eq:initialvalueproblem}
\left\{ \begin{array}{ll}
\dot{{\bf z}}(t) = \mathbf{f}(t,{\bf z}_t),\\
{\bf z}(t) = \x_{t_k}(t-t_k), & t\in[t_k-h, t_k), \\
{\bf z}(t_k) = \mathbf{I}(t_k,\mathbf{x}_{t_k}),
\end{array} \right.
\end{equation}
where $t\geq t_k$ and  $\x_{t_k} \in \mathcal{PC}$ is the initial condition. It will be assumed that the initial value problem (\ref{eq:initialvalueproblem}) has a unique solution for $t \geq t_k$ (see Corollary 3.1 in \cite{BL2000}). For example, that there exist constants $M$, $N \in \mathds{R}^+$ such that $\|\mathbf{f}(t,\psi)\| \leq M+N\|\psi\|$ for all $(t,\psi) \in \mathds{R}^+ \times \mathcal{PC}$;  \textcolor{black}{and that $\mathbf{f}$ is {\em locally Lipschitz}, that is for each compact set $\Omega \subset \mathds{R}^n$ there exists some constant $K \in \mathds{R}^+$ such that $\|\mathbf{f}(t,\psi_1) - \mathbf{f}(t,\psi_2) \| \leq K \|\psi_1 - \psi_2\|$ for all $t\in \mathds{R}^+$ and all $\psi_1, \psi_2 \in \mathcal{PC} ([-h,0],\Omega)$.}
%in particular it will be assumed that $\mathbf{f}$ is bounded,  that is
%$
%\|\mathbf{f}(t,\psi)\| \leq L
%$, 
%for some constant $L \in \mathds{R}^+$ and any $(t,\psi) \in [t_0, \infty)\times \mathcal{PC}$, and that $\mathbf{f}(t,\psi)$ is Lipschitz continuous in $\psi$, that is there exists a  constant $K\in\mathds{R}^+$ such that $\|\mathbf{f}(t,\psi_1) - \mathbf{f}(t,\psi_2) \| \leq K \|\psi_1 - \psi_2\|$ for any $\psi_1, \psi_2 \in \cal{PC}$ and any $t \in [t_0, \infty)$. 
Thus, the initial-value problem is well-posed and for every $(t_k,\x_{t_k}) \in \mathds{R}^+ \times \mathcal{PC}$, there exists a continuous and unique solution ${\bf z}(t_k,\x_{t_k})(t)$ for all $t \in [t_k,\infty)$.\\

 Hence, the solution $\x(t_0,\phi)$ of (\ref{eq:resetsystem}%-\ref{eq:resetset}) 
 ) is made up of an initial condition $\phi$ and a sequence of continuous solution segments ${\bf z}(t_k,\x_{t_k})$, that is 
\begin{equation}
\x(t_0,\phi)(t)=\left\{  \begin{array}{ll}
{\bf z}(t_0,\phi)(t), & t\in [t_0,t_1], \\
{\bf z}(t_1,\x_{t_1})(t), & t\in (t_1,t_2], \\
\multicolumn{2}{c}{\vdots} \\
{\bf z}(t_k,\x_{t_k})(t), & t\in (t_{k},t_{k+1}]. \\
\end{array} \right.
\end{equation}
If the sequence of reset instants is finite then $\x(t_0,\phi)(t)={\bf z}(t_N,\x_{t_N})(t)$, $t\in (t_{N},\infty)$. In addition, note that the solution is left-continuous with right limits, and there exists jump discontinuities at the reset instants $t_k$, $k = 1, 2, \cdots$, %if a reset instant is performed at the instant $t_k$, then the solution has a jump discontinuity at that instant,
that is the limits $\x(t_k^+)$ and $\x(t_k^-)$ exist, and $\x(t_k)=\x(t_k^-)$.\\

Suppose $\mathbf{f}(t,0)=0$ and $\mathbf{I}(t,0)=0$ for all $t\in \mathds{R}^+$. The trivial solution $\x_t=0$ of system (\ref{eq:resetsystem}%-\ref{eq:resetset}
) (henceforth named {\em zero solution}) is said to be {\em stable} if for any $t_0\in \mathds{R}^+$ and $\epsilon>0$, there exists $\delta = \delta(t_0, \epsilon)$ 
such that $\| \phi \| < \delta$ implies $ \| \x(t)\| < \epsilon$ for $t \geq t_0$. In addition, the solution $\x_t=0$ is {\em uniformly stable} if $\delta=\delta(\epsilon)$.  On the other hand, the zero solution is said to be {\em asymptotically stable} if it is stable and there exists $\delta_0=\delta_0(t_0) > 0$ such that $\lim_{t \rightarrow \infty} \| x(t)\| = 0$  whenever $\| \phi \|< \delta_0$. The solution is {\em uniformly asymptotically stable} if it is uniformly stable and there exists $\delta_0>0$ such that, for every $\eta>0$ there exists a $T=T(\eta)$ such that $\| \phi \| < \delta_0$ implies $\| \x(t) \| < \eta$ for $t \geq t_0 + T$ and for every $t_0\in \mathds{R}^+$; moreover, if $\delta_0$ can be an arbitrarily large finite number, then $\x_t=0$ is said to be {\em globally uniformly asymptotically stable}.\\

A function $f:\mathds{R}^+ \rightarrow \mathds{R}^+$ is said to be nondecreasing if $f(b)\geq f(a)$ for all $b>a$, where $a$, $b\in \mathds{R}^+$, if $f(b)> f(a)$ then it is said to be strictly increasing. In addition, $f$ is of class $\mathcal{K}$ if it is continuous, strictly increasing, and $f(0)=0$.\\

Let $V:  \mathds{R}^+ \times {\cal PC} \rightarrow \mathds{R^+}$ be continuously differentiable with respect to all of its arguments, and let $\mathbf{x}(t_0,\phi)(t)$ be the solution of the system (\ref{eq:resetsystem}%\ref{eq:resetset})
). Thus, $V(t,\x_t)$ only has (jump) discontinuities at $t\in \mathds{T}_\phi$. In addition, the upper right-hand derivative of $V$ along the solution $\mathbf{x}(t_0,\phi)$ is defined by
\begin{equation}
\dot{V}(t,\x_t)=\limsup_{\substack{
   \epsilon \rightarrow 0,\:
   \epsilon >0
  }} \frac{V(t+\epsilon,{ \bf x}_{t+\epsilon}) - V(t,{ \bf x}_t)}{\epsilon}
\label{eq:}
\end{equation}
for all  $t\in[t_0, \infty) \setminus \mathds{T}_\phi$. In addition, the increment of $V$ along the solution $\mathbf{x}(t_0,\phi)$ is defined by
\begin{equation}
%\Delta V(t,\psi) = V(t,\mathbf{I}(t,\psi)) - V(t,\psi)
\Delta V(t,\mathbf{x}_t) = V(t,\mathbf{I}(t,\mathbf{x}_t)) - V(t,\mathbf{x}_t)
\end{equation}
for any $t \in \mathds{T}_\phi$.%$(t,\psi) \in [t_0,\infty) \times {\cal PC}$

\section{Main Result}
\label{section:mainresult}

In this section, sufficient conditions for stability of the reset system (1) are proposed as a Lyapunov-Krasovskii theorem, generalizing the basic result for retarded functional differential equations (without reset actions) (see for example \cite{hale,gu}). \textcolor{black}{The result takes inspiration from \cite{hale}}; in fact, since in general, for a given initial condition $\phi$, the system (\ref{eq:resetsystem}) may not have reset actions and thus $\mathds{T}_\phi =\varnothing$, the proof is identical in that case. 
 \\

%{\em Remark 3.1}: The proof of Proposition 3.1 is strongly based on the Lyapunov-Krasowskii theorem for retarded functional differential equations (with no reset actions) as given in \cite{hale}; in fact, since in general, for a given initial condition $\phi$, the system (\ref{eq:resetsystem}) may not have  reset actions and thus $\mathds{T}_\phi = \emptyset$, the proof is identical in that case. 

%we establish Lyapunov-Krasovskii sufficient conditions for the (global) uniform (asymptotic) stability analysis of the system (\ref{eq:resetsystem}).\\

{\bf Proposition 3.1}: Assume that \textcolor{black}{$\mathbf{f}$ is locally Lipschitz}, $u, v, w:\mathds{R}^+ \rightarrow \mathds{R}^+$ are continuous nondecreasing functions and in addition $u,v \in \mathcal{K}$. If there exists a (Lyapunov-Krasovskii) functional $V:\mathds{R}^+ \times {\cal PC} \rightarrow \mathds{R^+}$ such that 
\begin{equation}\label{eq:cond1}
u(\|\psi(0)\|) \leq V(t,\psi) \leq v(\|\psi\|)
\end{equation}
\noindent for any $\psi \in \Omega= \{ \psi\in {\cal PC}:  \|\psi\| < \gamma\}$ for some $\gamma>0$ and all $t \in \mathds{R^+}$, and that for every solution $\mathbf{x}(t_0,\phi)$ of the system (\ref{eq:resetsystem}), $V(t,\mathbf{x}_t)$ is continuous for all $t\geq t_0$ and $t_0 \in \mathds{R}^+$ except on the set $\mathds{T}_\phi$, and in addition 
\begin{equation}\label{eq:cond2}
\dot{V}(t,\x_t) \leq -w(\|\x_t(0)\|), \hspace{1cm}  \mathbf{\x}_t(0) \notin \cal{M},
\end{equation}
\begin{equation}\label{eq:cond3}
\Delta V(t,\x_t) \leq 0, \hspace{1cm}  \mathbf{\x}_t(0) \in \cal{M},
\end{equation}
\noindent where $\dot{V}$ and $\Delta V$ are evaluated along the trajectories of (\ref{eq:resetsystem}) with $\x_t \in \Omega$, then the zero solution of (\ref{eq:resetsystem}) is uniformly stable. If $w(s) > 0$ for $s>0$ then the solution is uniformly asymptotically stable. In addition, if $\lim_{s \rightarrow \infty} u(s) = \infty$ and $\Omega={\cal PC}$, then it is globally uniformly asymptotically stable.\\

{\bf Proof}: \textcolor{black}{({\em Uniform stability)}} For a given $\epsilon  > 0$, let set $\epsilon_1<\min(\epsilon, \gamma)$, then it can be found some $\delta=\delta(\epsilon_1)$, $0<\delta < \epsilon_1$, such that $v(\delta) < u(\epsilon_1)$. Suppose that $\x(t_0,\phi)(t)$ is the solution of (\ref{eq:resetsystem}) for $(t_0,\phi)\in \mathds{R}^+ \times {\cal PC}$%, where $\|\phi \| < \delta$
 . Therefore, $\x(t_0,\phi)(t)$ is continuous on $[t_0-h,\infty)\setminus \mathds{T}_\phi$, where $\mathds{T}_\phi = \{t_1, t_2, \cdots\}$ is the set of reset instants corresponding to the initial condition $\phi$, and $t_0 <  t_1$. Now, we will prove that $\|\x(t_0,\phi)(t)\| < \epsilon_1 < \epsilon$ for \textcolor{black}{any initial condition $\phi$, with $\|\phi \| < \delta$, and $t\geq t_0$. By contradiction, if it is false then $\|\x(t_0,\phi^\star)(t^\star)\| \geq \epsilon_1$ at some instant $t ^\star\geq t_0$, and for some initial condition $\phi^\star$, with $\|\phi^\star \| < \delta$
 %Suppose that  the previous inequality is violated 
. Let $T\geq t_0$ be given by $T = \min\{t^\star \in \mathds{R}^+: \lim_{ s \rightarrow t^\star , s>t^\star} \|\x(t_0,\phi^\star)(t^\star)\| \geq \epsilon_1\}$. 
%the smallest value such that $\lim_{ s \rightarrow T , s>T} \| \x(s) \| \geq \epsilon_1$. This quantity is well-defined since the solution $\x(t_0,\phi)(t)$ is left continuous. 
Note that if 
$T\in \mathds{T}_{\phi^\star}$ then $\| \mathbf{I}(T,\x_T) \| = \epsilon_1$, otherwise $\| \x(T) \| = \epsilon_1$; thus, as a result, we have $\| \x(t_0,\phi^\star)(t) \| < \epsilon_1$ for $t\in [t_0,T)$. In addition, let $\mathds{T}_{\phi^\star} = \{t_1^\star, t_2^\star, \cdots\}$ be the sequence of reset instants corresponding to the initial condition $\phi^\star$; since reset instants are well-posed and Zeno-free then there exists $N>0$, defined as the largest integer for which $t_N^\star \leq T$. 
 Now, since $\delta<\epsilon_1$, conditions (\ref{eq:cond2}) and (\ref{eq:cond3}) imply
\begin{equation}
V(t,\mathbf{x}_t) \leq V(t_k^\star, I(t_k^\star ,\mathbf{x}_{t_k^\star})) \leq V(t_k^\star,\mathbf{x}_{t_k^\star})
\end{equation}
\noindent for $t \in (t_k^\star, t_{k+1}^\star]$, $k= 1, 2, \cdots, N-1$, and 
\begin{equation}
V(t,\mathbf{x}_t) \leq  V(t_0,\phi^\star)
\end{equation}
for $t \in [t_0, t_1^\star]$. Since, in addition, $V(T,\mathbf{x}_T) \leq V(t_N^\star,\mathbf{x}_{t_N\star})$, and 
$\|\phi^\star \| < \delta < \epsilon_1 \leq \gamma$, combining (9)-(10) and (\ref{eq:cond1}) it follows}
\begin{equation}
V(T,\mathbf{x}_T) \leq V(t_0,\phi^\star) \leq v(\|\phi^\star\|) < v(\delta) < u(\epsilon_1), 
\end{equation}
But, from (\ref{eq:cond1}) and (\ref{eq:cond3}), it follows
\begin{equation}
u(\epsilon_1)=u(\| \x(T) \| )  \leq V(T,\x_T) \ \ \text{if} \ \ T\notin \mathds{T}_{\phi^\star}
\end{equation}
and
\begin{equation}
u(\epsilon_1) = u(\| \mathbf{I}(T,\x_T) \| ) \leq V(T,\mathbf{I}(T,\x_T)) \leq V(T,\x_T)\ \ \text{if} \ \ T \in \mathds{T}_{\phi^\star},
\end{equation}
which is a contradiction in both cases. %Therefore, the original assumption is false, and $\| \x(t) \| < \epsilon_1 < \epsilon$ for $t\geq t_0$. 

%Finally, since $\|\phi \| < \delta<\epsilon$, it is proved that the zero solution is uniformly stable.
 
\textcolor{black}{({\em Uniform asymptotic stability)}} In this case, the proof  is a bit more involved. For $\epsilon > 0$ choose  $\delta_a > 0$ such as \textcolor{black}{$v(\delta_a) < u(\min\{\epsilon,\gamma\})$, }thus it is true that $\|\phi\| < \delta_a$ implies $\|\mathbf{x}(t_0,\phi)(t)\| < \min(\epsilon,\gamma)$ for $t \geq t_0$. Now, it will be shown that for any $\eta > 0$ there exists some $T(\delta_a,\eta)$ such that $\|\mathbf{x}(t_0,\phi)(t)\| < \eta$ \textcolor{black}{for any $\phi$, with $\|\phi\| < \delta_a$, and} $t \geq t_0 + T$.  This is equivalent to prove that $\|\mathbf{x}_{t_0+T}\| < \delta_b$, where \textcolor{black}{$v(\delta_b) =u(\min\{\eta,\gamma\})$.} %the constant for uniform stability. 
By contradiction, suppose that there not exists such $T$, \textcolor{black}{that is there exist some $\eta>0$ and a solution $\mathbf{x}(t_0,\phi^\star)(t)$, with $\| \phi^\star \| < \delta_a$,} such as $\|\mathbf{x}_{t}\| \geq \delta_b$ for all $t \geq t_0$. Thus, there exists a sequence $(\tau_k)$, $k = 1, 2, \cdots$ such that 
\begin{equation}
t_0 + (2k-1)h \leq \tau_k \leq t_0 + 2kh,
\end{equation}
\noindent where $\tau_k \notin \mathds{T}_{\phi^\star} =  \{t_1^\star, t_2^\star, \cdots\}$ and $\|\x(\tau_k)\|\geq\delta_b$. \textcolor{black}{Since it is assumed that the system (\ref{eq:resetsystem}) has well-posed and Zeno-free reset instants, and $t_0$ is not a reset instant, then $t_0 < t_{1}^\star < t_2^\star < \cdots $, and if $\mathds{T}_{\phi^\star}$ is infinite then $t_N^\star \rightarrow \infty$ as $N\rightarrow \infty$. In addition, since ${\bf f}$ is locally Lipschitz, and by uniform stability $\|\mathbf{x}(t_0,\phi^\star)(t)\| < \min(\epsilon,\gamma)$ for $t \geq t_0$, then there exists a constant $L>0$ such that $\|\dot{\mathbf{x}}(t)\| < L$ for all $t \in [t_0,\infty)\setminus \mathds{T}_\phi$.} Therefore, it is possible to build a set of intervals $I_k = [\tau_k - \frac{\delta_b}{2L}\alpha^1_k , \tau_k + \frac{\delta_b}{2L}\alpha^2_k] $ with $\alpha^1_k$, $\alpha^2_k \in \{0,1\}$, $\alpha^1_k+\alpha^2_k\geq1$, that do not contain reset instants and do not overlap (by using a number $L > 0$ large enough and proper values of $\alpha^1_k$, $\alpha^2_k$), that is \textcolor{black}{$I_k \cap \mathds{T}_{\phi^\star} = \varnothing$}, $k = 1, 2, \cdots$, and then by using the mean-value theorem on the intervals $[\tau_k,t] \subset I_k$ and $[t,\tau_k] \subset I_k$
\textcolor{black}{
\begin{equation}
\|\mathbf{x}(t_0,\phi^\star)(t)\| = \| \mathbf{x}(\tau_k) + \dot{\mathbf{x}}(\tau_k + \theta_1 (t-\tau_k))(t-\tau_k)\|,
\end{equation}
\begin{equation}
\|\mathbf{x}(t_0,\phi^\star)(t)\| = \| \mathbf{x}(\tau_k) - \dot{\mathbf{x}}(\tau_k + \theta_2 (\tau_k-t))(\tau_k-t)\|
\end{equation}
}
for some $\theta_1,\theta_2 \in (0,1)$, and then
\textcolor{black}{
\begin{equation}
\|\mathbf{x}(t_0,\phi^\star)(t)\| \geq  \| \mathbf{x}(\tau_k)\|  - \|\dot{\mathbf{x}}(\tau_k + \theta (t-\tau_k))\| |t-\tau_k| \geq \frac{\delta_b}{2}
\end{equation}
}
for any $t \in I_k$ and some $\theta \in (-1,1)$. In addition, from (\ref{eq:cond2}) it is true that $\dot{V}(t,\mathbf{x}_t) \leq -w(\|\mathbf{x}(t)\|) \leq - w(\frac{\delta_b}{2})< 0$, for any $t \in I_k$, this means that $V(t,\mathbf{x}_t)$ is decreasing with at least a ratio $-w(\frac{\delta_b}{2})$ in each interval $I_k$, $k= 1,2, \cdots$. On the other hand, the reset instants in the sequence \textcolor{black}{$\mathds{T}_{\phi^\star} = \{t_1^\star,t_2^\star, \cdots \}$ may  be renamed as  $\mathds{T}_{\phi^\star} = \{t_{k,l}^\star\}$, where the reset instant $t_{k,l}^\star$ corresponds to the $l^{th}$-instant prior to $\tau_k \in I_k$, that is 
\begin{equation}
\begin{array}{l}
t_0 \leq t_{1,1}^\star <  t_{1,2}^\star < \cdots t_{1,N_1}^\star < \check{\tau}_{1} = \tau_{1} - \frac{\delta_b}{2L}\alpha^1_k  \\ < \hat{\tau}_{1} = \tau_{1} + \frac{\delta_b}{2L}\alpha^2_k < t_{2,1}^\star  < \cdots t_{2,N_2}^\star <  \check{\tau}_{2} < \cdots
\end{array}
\end{equation}
\noindent for some integers $N_1, N_2, \cdots \geq 0$. Note that $N_1, N_2, \cdots$ do exist since reset instants are well-posed and Zeno-free.} Therefore, by integrating $\dot{V}(t,\mathbf{x}_t)$ over the interval $[\check{\tau}_1, \hat{\tau}_1]$ and using also (\ref{eq:cond3}), it is obtained
\textcolor{black}{
\begin{equation}
\begin{array}{l}
V(\check{\tau}_{1} ,\mathbf{x}_{\check{\tau}_{1}} ) \leq V(t_{1,N_1}^\star,\mathbf{I}(t_{1,N_1}^\star,\mathbf{x}_{t_{1,N_1}^\star}))  \leq  V(t_{1,N_1}^\star,\mathbf{x}_{t_{1,N_1}^\star}) \leq \cdots \leq V(t_0,\phi^\star) 
\end{array}
\end{equation}
}
and
\begin{equation}
\begin{array}{l}
V(\tau_2,\mathbf{x}_{\tau_2}) \leq V(\hat{\tau}_{1} ,\mathbf{x}_{\hat{\tau}_{1} }) \leq V(\check{\tau}_{1} ,\mathbf{x}_{\check{\tau}_{1} }) - w(\frac{\delta_b}{2})\frac{\delta_b}{2L}(\alpha^1_k+\alpha^2_k) \\ \textcolor{black}{\leq \cdots \leq V(t_0,\phi^\star)- w(\frac{\delta_b}{2})\frac{\delta_b}{2L} (\alpha^1_k+\alpha^2_k).}
\end{array}
\end{equation}  
\noindent Finally, since $\alpha^1_k+\alpha^2_k\geq 1$ for $k=1,2,\ldots$, then repeating the reasoning for any $\tau_k$, it follows
\textcolor{black}{
\begin{equation}
V(\tau_k,\mathbf{x}_{\tau_k}) \leq V(t_0,\phi^\star) - w(\frac{\delta_b}{2})(k-1)\frac{\delta_b}{2L},
\end{equation}
}
\noindent and then for a large enough $k$ it results that $V(\tau_k,\mathbf{x}_{\tau_k}) < 0$, which is a contradiction. 
Finally, if $\lim_{s \rightarrow \infty} u(s) = \infty$ and $\Omega=\cal{PC}$, then $\delta_a$ above may be chosen arbitrarily large, and $\epsilon$ can be set after $\delta_a$ to satisfy $v(\delta_a) < u(\epsilon)$. Therefore, global uniform asymptotic stability can be concluded. $\Box$ \\

\section{Application cases}
\label{section:application}

\textcolor{black}{In the following cases, it will be assumed that reset instants are time-regularized, and thus they are well-posed and Zeno-free. In addition, the rest of assumptions in Prop. 3.1 can be easily checked to be satisfied, and it will not be explicitly shown. }

\subsection{Reset systems with a LTI base system and single time-delay }

In this section, we establish delay-independent stability conditions for time-delay reset systems with LTI base system as in \cite{BanBar}, here the provided proof is linked to the main stability result of Section \ref{section:mainresult}. Consider a special case of (\ref{eq:resetsystem}), in which the base system is a LTI base system with a single time-delay, given by

\begin{equation}\label{eq:ltiresetsystem}
\left\{ \begin{array}{ll}
 \dot{\mathbf{x}}(t) =  A \x(t)+ A_d \x(t-h), &   \mathbf{x}(t) \notin {\cal M},   \\
{\mathbf{x}}(t^+) = A_R \x(t), &  \mathbf{x}(t) \in {\cal M},\\
\mathbf{x}(t) =  \phi(t),  & t \in [-h,0],
\end{array}\right.
\end{equation}
for arbitrary values of $A$ and $A_d$, and where $t_0=0$ is the initial instant, and the reset matrix $A_R$ takes the form $A_R=diag(I_{\bar{n}_\rho},0_{n_\rho})$, $\bar{n}_\rho = n - n_\rho $. The reset action is applied on the last $n_\rho$ states of the vector $\x\in \mathds{R}^n$ at those instants in which the state reaches the reset set defined as \textcolor{black}{$\mathcal{M} =  \{{\bf x} \in \mathds{R}^n: C\mathbf{x}=0 \}$ for some row vector $C \in \mathds{R}^{1 \times n}$}. %Note that the system (\ref{eq:ltiresetsystem}) is implicitly time regularized by some $\Delta>0$, being the interval between two consecutive reset instants lower bounded by $\Delta$. Hence, there exists a unique solution of the system, as in the nonlinear case. 
The asymptotic stability of this system can be analyzed by the following proposition.\\

{\bf Proposition 4.1}: If there exist (symmetric) matrices $P$, $Q>0$ such that 
\textcolor{black}{
\begin{equation}\label{eq:LTIcond1}
 M= \left(
\begin{array}{cc}
A^\top P + PA + Q   &   PA_d  \\
 A_d^\top P &  -Q
\end{array}
\right)<0
\end{equation}
}
and 
\begin{equation}\label{eq:LTIcond2}
\Theta^\top ( A_R^\top P A_R - P ) \Theta \leq 0
\end{equation}
for some $\Theta$ with $\mathcal{R} (\Theta)=\mathcal{N}(C)$, then the zero solution of system (\ref{eq:ltiresetsystem}) is globally asymptotically stable.\\

{\bf Proof}: Consider the Lyapunov-Krasowskii functional $V:{\cal PC} \rightarrow \mathds{R^+}$ given by
\begin{equation}\label{eq:functional}
 V(\psi)  = \psi(0)^\top P \psi(0) \ + \int_{-h}^{0}  \psi^\top(\theta) Q  \psi(\theta) \ d\theta,
\end{equation}
with $P$, $Q$ the matrices of the proposition. For this functional, since $P$, $Q > 0$ it is true that
\begin{equation} 
V(\psi) \leq \lambda_{M}(P) \|\psi(0)\|^2 + h\lambda_{M}(Q)\|\psi\|^2  \leq (\lambda_{M}(P)+ h\lambda_{M}(Q))\|\psi\|^2 = v(\|\psi\|)
\end{equation}
\noindent and
\begin{equation}
V(\psi) \geq \lambda_{m}(P) \|\psi(0)\|^2 + h\lambda_{m}(Q)\|\psi\|^2  \geq \lambda_{m}(P)\|\psi(0)\|^2 = u(\|\psi(0)\|),
\end{equation}
\noindent where $u,v: \mathds{R}^+ \rightarrow \mathds{R}^+$ are continuous nondecreasing functions and $u,v \in K$. On the other hand, the derivative of $V$ along the solutions of (\ref{eq:ltiresetsystem}), after some manipulation, is given by 
\begin{equation}
\dot{V}(\psi) = 
\left(
\begin{array}{cc}
  \x^\top(t)& \x^\top(t-h)     
\end{array}
\right)
M\left(
\begin{array}{c}
  \x(t)  \\ \x(t-h)
\end{array}
\right).
\end{equation}
Therefore, condition (\ref{eq:LTIcond1}) is obtained, and it implies
\begin{equation}
- \dot{V}(\x_t) > \lambda_{m}(-M)(\|\x(t)\|^2 + \|\x(t-h)\|^2)   \geq \lambda_{m}(-M)(\|\x(t)\|^2) = w(\|\x(t)\|)
\end{equation}
\noindent for all $\x(t) \notin {\cal M}$, where $w: \mathds{R}^+ \rightarrow \mathds{R}^+$ is a continuous nondecreasing function and $w \in K$. Finally, condition (\ref{eq:LTIcond2}) is obtained by setting $\Delta V(\x_t)= \x^\top(t) (A_R^\top P A_R -P) \x(t)\leq0$ and considering that $\x \in \cal{M}$ implies that there exist $y$ such that $x=\Theta y$. As a result, all the conditions of Proposition 3.1 are satisfied and thus the zero solution of the system (\ref{eq:ltiresetsystem}) is globally asymptotically stable. $\Box$ \\

\textcolor{black}{
{\bf Example 4.1:} Consider the time-delay reset system (\ref{eq:ltiresetsystem})  with matrices
\begin{equation}
\begin{array}{ccccc}
A=\left( \begin{array}{cc} -2 & 0 \\ 0 & -0.9
\end{array}\right), & &
A_d=\left( \begin{array}{cc} -1 & 1 \\ -1 & -0.5
\end{array}\right), & &
A_R=\left( \begin{array}{cc} a_1 & 0 \\ a_2 & a_3
\end{array}\right).
\end{array}
\end{equation}
The base system is asymptotically stable independently of the time-delay since there exist matrices $P$ and $Q$ such that condition (\ref{eq:LTIcond1}) is satisfied (see for example \cite{gu}). Now suppose that $C= (\ 1 \ \ 0 \ )$, then condition (\ref{eq:LTIcond2}) is satisfied whenever $| a_3 | \leq 1$, and thus, the reset system is globally asymptotically stable. Moreover, if $a_2=0$,  $| a_1 | \leq 1$ and $| a_3 | \leq 1$ then the asymptotic stability of the reset control system is guaranteed for any row vector $C$. The trajectory of the reset system with $C= [\ -2 \ \ 1 \ ]$, $h=1$, $a_1=1$, $a_2=0$ and $a_3=0$ is shown in Fig. \ref{fig:LTI_trajectory}. In addition, Fig. \ref{fig:LTI_functional} shows the value of the Lyapunov-Krasovskii functional along this trajectory. Note that the Lyapunov-Krasovskii functional obeys the two conditions in Prop. 4.1, decreasing both during the continuous dynamic and the jumps. 
\begin{figure}
\begin{minipage}[b]{.5\textwidth}
  \vspace*{\fill}
  \centering
 \includegraphics[scale=1]{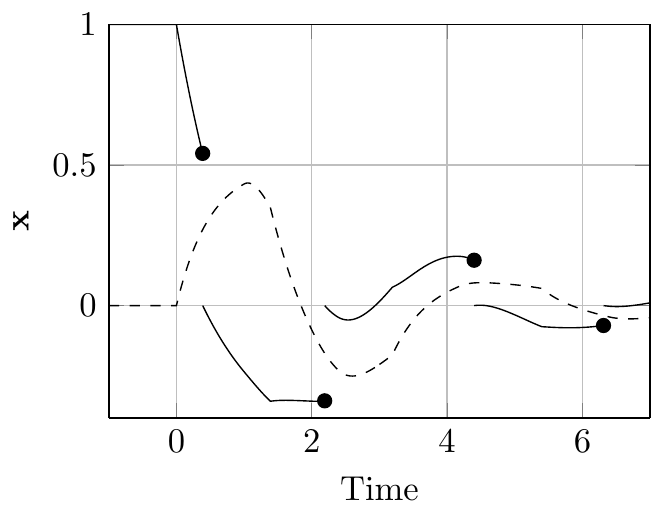}
        \subcaption{\textcolor{black}{Trajectory:  $x_1$ (dashed) and $x_2$ (solid).} }
        \label{fig:LTI_trajectory}
\end{minipage}
\begin{minipage}[b]{.5\textwidth}
  \vspace*{\fill}
  \centering
 \includegraphics[scale=1]{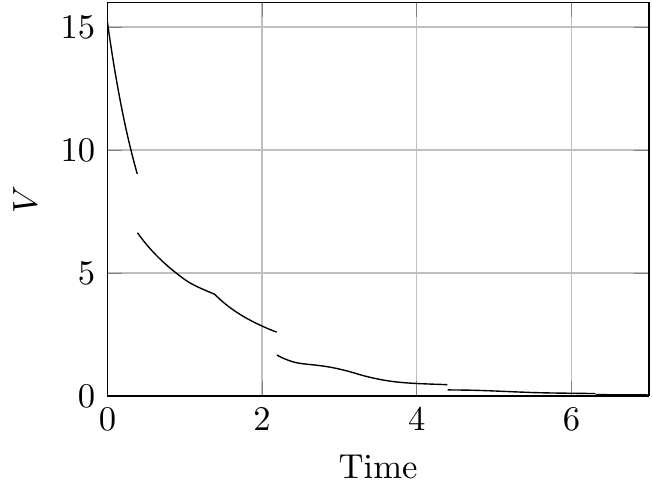}
        \subcaption{\textcolor{black}{Lyapunov-Krasovskii functional.}}
        \label{fig:LTI_functional}
\end{minipage}
  \caption{\textcolor{black}{Trajectory and value of the Lyapunov-Krasovskii functional in the Example 4.1.}}
   \label{fig:LTIexample}
\end{figure}
}

\subsection{Reset system with a nonlinear and time varying base system }

In general, for a reset system with nonlinear and time-varying base system without a particular structure, there is not systematic procedure for generating Lyapunov-Krasovskii functionals candidates. Therefore, in this section the main result is applied to a reset system with a particular structure.\\

{\bf Example 4.2:} Consider a second-order reset system given by
\begin{equation}\label{eq:nonlinearresetsystem}
\left\lbrace \begin{array}{ll}
\left\lbrace \begin{array}{l}
											\dot{x}_1(t) = a_1(t) x_1^3 (t) +  b_1(t) x_1^3 (t-h) + c_1(t) x_2(t) \\
											\dot{x}_2(t) = a_2(t) x_2 (t) +  b_2(t) x_2 (t-h) + c_2(t) x_1^3(t)									
							\end{array}	
					\right\rbrace,& \textcolor{black}{{\bf g}(x_1(t),x_2(t))\neq 0, 	}
					\\
					\\
\left\lbrace \begin{array}{l}
											x_1(t^+) = x_1(t) \\
											x_2(t^+) = 0									
							\end{array}	
\right\rbrace,& \textcolor{black}{{\bf g}(x_1(t),x_2(t))= 0. }		
\end{array}		\right.				
\end{equation}
where $a_1(t)$, $b_1(t)$, $c_1(t)$, $a_2(t)$, $b_2(t)$, and $c_2(t)$ are arbitrary continuous bounded functions with $a_1(t), a_2(t) \leq -\delta$, $|b_1(t)|,|b_2(t)| \leq \frac{\delta}{2}$, and $|c_1(t)|,|c_2(t)| \leq \frac{\delta}{4}$ for some given $\delta>0$. \textcolor{black}{Here, the reset set $\mathcal{M}$ is given as  $\mathcal{M} = \{(x_1,x_2) \in \mathds{R}^2: \mathbf{g}(x_1,x_2)= 0, \text{for some } \mathbf{g}: \mathds{R}^2  \rightarrow \mathds{R} \}$.	}

Consider the candidate functional $V$ be defined by 
\textcolor{black}{
\begin{equation}
\label{eq:non_functional}
V(\x_t) = \frac{x_1^4(t)}{4}+ \frac{x_2^2(t)}{2} + \frac{\delta}{2}\int^0_{-h} \left( x_1^6(t+\alpha) +  x_2^2(t+\alpha) \right) d\alpha,
\end{equation}
}
where $\x(t)=(x_1(t),x_2(t))\in\mathds{R}^2$ and $\x_t=x(t+\alpha)$, $\alpha \in [-h,0]$. Let define the continuous nondecreasing functions $u(s)=\frac{1}{4}s^4$ and $v(s)=2 s^2$, then the above functional satisfies condition (\ref{eq:cond1}),
\begin{equation}
u(\|\x_t(0)\|)=\frac{1}{4}(x_1^2(t)+x_2^2(t))^2 \leq \frac{1}{4}(x_1^4(t)+x_2^2(t)) \leq V(\x_t)
\end{equation}
and
\begin{equation}
V(\x_t)\leq  x_1^2(t)+ x_2^2(t)  + \frac{\delta}{2}\int^0_{-h} \left( x_1^2(t+\alpha)+ x_2^2(t+\alpha)\right) d\alpha \leq 2\|\x_t\|^2.
\end{equation}
It is easy to check that the above conditions are satisfied for any $\|\x_t\| \leq \gamma=\sqrt{\frac{1}{5} }$.

On the other hand, the derivative of the functional along the solutions of system (\ref{eq:nonlinearresetsystem}) is given by
\begin{equation}
\dot{V}(\x_t) = x_1^3(t)\dot{x}_1(t)+ x_2(t)\dot{x}_2(t) + \frac{\delta}{2}(x_1^6(t) - x_1^6(t-h))  +  \frac{\delta}{2}(x_2^2(t) - x_2^2(t-h)).
\end{equation}
After some manipulations the derivative of the functional is bounded by
\begin{equation}
\dot{V}(\x_t)\leq \xi^\top(t) M \xi(t),
\end{equation}
where $\xi(t)=\left( |x_1(t)|^3, |x_1(t-h)|^3,  |x_2(t)|,   |x_2(t-h)|  \right)$ and 
\begin{equation}
M=\left( \begin{array}{rrrr}
-\frac{\delta}{2} & \frac{\delta}{4} & \frac{\delta}{4} & 0\\
\frac{\delta}{4} & -\frac{\delta}{2} & 0 & 0\\
\frac{\delta}{4} & 0 & -\frac{\delta}{2} & \frac{\delta}{4}\\
0& 0 & \frac{\delta}{4} & -\frac{\delta}{2}\\
\end{array}\right).
\end{equation}
Since $M<0$ for any $\delta>0$, then defining $w(s)=\lambda_{m}(-M)s^6$ it follows
\begin{equation}
-\dot{V}(\x_t)\geq \lambda_{m}(-M) \left( |x_1(t)|^6 + |x_1(t-h)|^6+|x_2(t)|^2 \right.  \left. +|x_2(t-h)|^2 \right)\geq w(\| \x(t)\|)
\end{equation}
for any $\| \x_t\| \leq \sqrt{\frac{1}{5}}$, and thus condition (\ref{eq:cond2}) is satisfied.  Finally, $\Delta V(\x_t)=-\frac{x^2_2(t)}{2}$, which is negative for any reset instant, \textcolor{black}{ regardless of the function ${\bf g}$.} Hence, the solution $\x_t=0$ of the system (\ref{eq:nonlinearresetsystem}) is uniformly asymptotically stable.

\begin{figure}
\begin{minipage}[b]{.5\textwidth}
  \vspace*{\fill}
  \centering
 \includegraphics[scale=1]{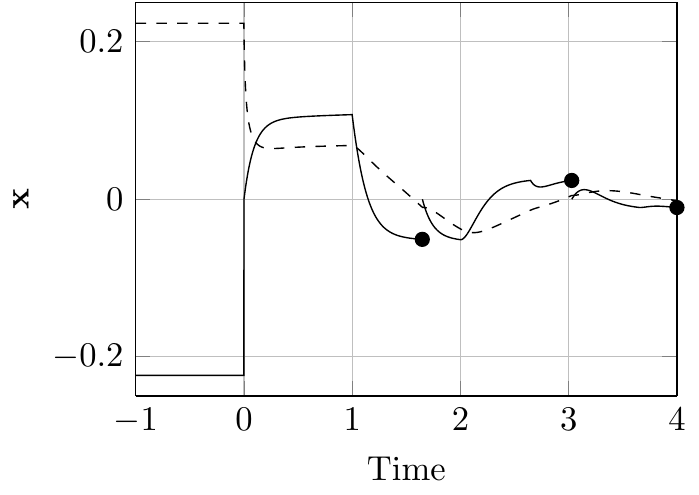}
        \subcaption{\textcolor{black}{Trajectory:  $x_1$ (dashed) and $x_2$ (solid). }}
        \label{fig:non_trajectory}
\end{minipage}
\begin{minipage}[b]{.5\textwidth}
  \vspace*{\fill}
  \centering
 \includegraphics[scale=1]{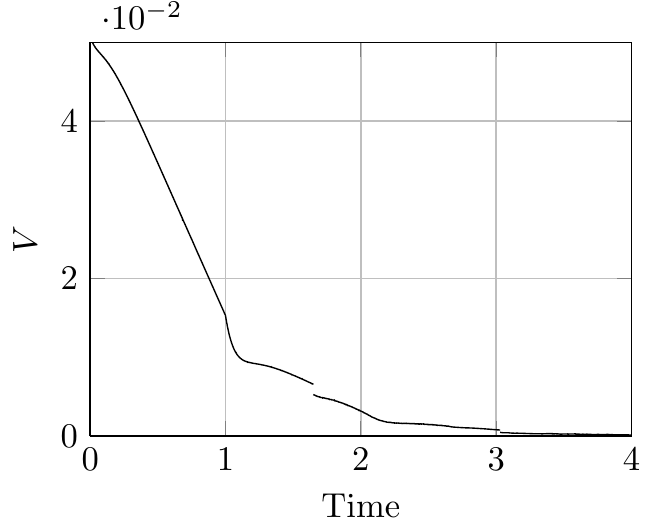}
        \subcaption{\textcolor{black}{Lyapunov-Krasovskii functional.}}
        \label{fig:non_functional}
\end{minipage}
  \caption{\textcolor{black}{Trajectory and value of the Lyapunov-Krasovskii functional in the Example 4.2.}}
   \label{fig:nonexample}
\end{figure}

\textcolor{black}{Now, consider the reset system with functions $a_1(t)$, $b_1(t)$, $c_1(t)$, $a_2(t)$, $b_2(t)$, and $c_2(t)$ given by
\begin{equation}
\begin{array}{lllll}
a_1(t)=50e^{-t}-100\delta, & & b_1(t) =\frac{\delta}{2}\sin(t), & &c_1(t)=\frac{\delta}{4},\\
a_2(t)=-e^{-t}-\delta, & &b_2(t)  =-\frac{\delta}{2}, & &c_2(t) =\frac{\delta}{4}\sin(t).
\end{array}
\end{equation}
It can be easily seen that the above functions satisfy the required bounds for any $\delta>0.51$. In addition, consider the function ${\bf g}(x_1,x_2)=-5x_1+x_2$ and the initial condition $\phi(t)=(\frac{1}{2\sqrt{5}}, -\frac{1}{2\sqrt{5}})$, $t \in [t_0-h,0]$ such that $\| \phi \| = \frac{1}{\sqrt{10} }< \frac{1}{\sqrt{5}}$. The evolution of the system with $h=1$ and $t_0=0$ is plotted in Fig. \ref{fig:non_trajectory}. In addition, Fig. \ref{fig:non_functional} shows the value of the functional (\ref{eq:non_functional}) along the trajectory. Once again, it can be observed how the functional always decreases.}

\section{ Conclusions}
\label{section:conclusion}
This work provides Lyapunov-Krasovskii based conditions that guarantee the stability of time-delay reset systems. In comparison with previous works, the main contribution has been to consider reset systems, whose base system is nonlinear and time-varying. As an application, the proposed result have been applied to obtain delay-independent condition in terms of LMI for time-delay reset control systems. In addition, the stability of a particular reset system with nonlinear and time-varying base system is analyzed.

\section*{Acknowledgements}
This work has been supported by Ministerio de Econom{\'i}a e Innovaci{\'o}n of Spain under project DPI2013-47100-C2-1-P (including FEDER co-funding).
The authors thank helpful comments of Sophie Tarbouriech and Luca Zaccarian, that have motivated   the problem approached in this work.

\section*{References}
\bibliography{referencias}

\begin{thebibliography}{10}
\expandafter\ifx\csname url\endcsname\relax
  \def\url#1{\texttt{#1}}\fi
\expandafter\ifx\csname urlprefix\endcsname\relax\def\urlprefix{URL }\fi
\expandafter\ifx\csname href\endcsname\relax
  \def\href#1#2{#2} \def\path#1{#1}\fi

\bibitem{clegg}
J.~C. Clegg, A nonlinear integrator for servomechanisms, Trans. of the American
  Institute of Electrical Engineering 77 (1958) 41--42.

\bibitem{KriHor}
K.~R. Krishman, I.~M. Horowitz, Synthesis of a non-linear feedback system with
  significant plant-ignorance for prescribed system tolerances, International
  Journal of Control 19~(4) (1974) 689--706.

\bibitem{HorRos}
I.~M. Horowitz, P.~Rosenbaum, Non-linear design for cost of feedback reduction
  in systems with large parameter uncertainty, International Journal of Control
  21~(6) (1975) 977--1001.

\bibitem{BanBarBook}
A.~Ba{\~n}os, A.~Barreiro, Reset Control Systems, Advances in Industrial
  Control, Springer, 2012.

\bibitem{astrom}
K.~J. \r{A}str\"om, Limitations on control system performance, European Journal
  of Control 6 (2000) 2--20.

\bibitem{beker2001}
O.~Beker, C.~Hollot, Y.~Chait, Plant with integrator: an example of reset
  control overcoming limitations of linear feedback, IEEE Transaction on
  Automatic Control 46~(11) (2001) 1797--1799.

\bibitem{davo}
M.~A. Dav{\'o}, A.~Ba{\~n}os, Reset control of a liquid level process, in: IEEE
  18th Conference on Emerging Technologies \& Factory Autom., IEEE, 2013, pp.
  1--4.

\bibitem{moreno}
J.~Moreno, J.~Guzm{\'a}n, J.~Normey-Rico, A.~Ba{\~n}os, M.~Berenguel, A
  combined {FSP} and reset control approach to improve the set-point tracking
  task of dead-time processes, Control Engineering Practice 21~(4) (2013)
  351--359.

\bibitem{BekHoll}
O.~Beker, C.~V. Hollot, Y.~Chait, H.~Han, Fundamental properties of reset
  control systems, Automatica 40~(6) (2004) 905--915.

\bibitem{hale}
J.~K. Hale, S.~M.~V. Lunel, Introduction to functional differential equations,
  Springer, New York, 1993.

\bibitem{gu}
K.~Gu, J.~Chen, V.~Kharitonov, Stability of Time-Delay Systems, Birkh{\"a}user,
  2003.

\bibitem{F2014}
E.~Fridman, Introduction to Time-Delay Systems, Systems \& Control: Foundations
  \& Applications, Birkh{\"a}user, 2014.

\bibitem{fridman}
E.~Fridman, A descriptor system approach to ${H}_\infty$ control of linear
  time-delay systems, {IEEE} Transaction on Automatic Control 47~(2) (2002)
  253--270.

\bibitem{Cao2004}
J.~Cao, J.~Wang, Delay-dependent robust stability of uncertain nonlinear
  systems with time-delay, Applied Mathematics and Computation 154~(1) (2004)
  289--297.

\bibitem{Cao2006}
F.~Ren, J.~Cao, Novel $\alpha$-stability criterion of linear systems with
  multiple time delays, Applied Mathematics and Computation 181~(1) (2006)
  282--290.

\bibitem{GP2006}
F.~Gouaisbaut, D.~Peaucelle, Delay-dependent stability of time-delay systems,
  in: Proceedings of the 5$^{th}$ IFAC Symposium on Robust Control Design,
  Toulouse, France, 2006, pp. 453--458.

\bibitem{SG2013}
A.~Seuret, F.~Gouaisbaut, Wirtinger-base integral inequality: application to
  time-delay systems, Automatica 49~(9) (2013) 2860--2866.

\bibitem{BanBar}
A.~Ba{\~n}os, A.~Barreiro, Delay-independent stability of reset control
  systems, {IEEE} Transaction on Automatic Control 54 (2009) 341--346.

\bibitem{BarBan}
A.~Barreiro, A.~Ba{\~n}os, Delay-dependent stability of reset systems,
  Automatica 46 (2010) 216--221.

\bibitem{BS1989}
D.~D. Bainov, P.~S. Simeonov, Systems with Impulse Effect: Stability, Theory
  and Applications, Ellis Horwood, Chichester, UK, 1989.

\bibitem{LBS1989}
V.~Lakshmikantham, D.~D. Bainov, P.~S. Simeonov, Theory of Impulsive
  Differential Equations, World Scientific, Singapore, 1989.

\bibitem{HCN2006}
W.~M. Haddad, V.~S. Chellaboina, S.~G. Nersesov, Impulsive and Hybrid Dynamical
  Systems: Stability, Dissipativity, and Control, Princeton Series in Applied
  Mathematics, 2006.

\bibitem{MHD2008}
A.~N. Michel, L.~Hou, D.~Liu, Stability of dynamical systems: continuous,
  discontinuous, and discrete systems, Birkhauser, Boston, 2007.

\bibitem{r_BM2012}
A.~Ba{\~n}os, J.~I. Mulero, Well-posedness of reset control systems as
  state-dependent impulsive dynamical sysetms, Abstract and Applied Analysis
  2012.

\bibitem{BL2000}
G.~Ballinger, X.~Liu, Existence, uniqueness and boundedness results for
  impulsive delay differential equations, Applicable Analysis: An international
  Journal 74~(1-2) (2000) 71--93.

\bibitem{delasen}
M.~de~la Sen, N.~Luo, A note on the stability of linear time-delay systems with
  impulsive inputs, IEEE Transactions on Circuits and Systems I: Fundamental
  Theory and Applications 50~(1) (2003) 149--152.

\bibitem{impulsivo2004}
X.~Liu, Stability of impulsive control systems with time delay, Mathematical
  and Computer Modelling 39 (2004) 511--519.

\bibitem{impulsivo2007-2}
X.~Liu, X.~Shen, Y.~Zhang, Q.~Wang, Stability criteria for impulsive systems
  with time delay and unstable system matrices, IEEE Transaction on circuits
  and systems 54~(10) (2007) 2288--2298.

\bibitem{impulsivo2008}
P.~Haghshtabrizi, J.~P. Hespanha, A.~R. Teel, Stability of delay impulsive
  systems with application to networked control systems, Transaction of the
  Institute of measurement and control 32~(5) (2009) 511--528.

\bibitem{SMZ2005}
Y.~Sun, A.~N. Michel, G.~Zhai, Stability of discontinuous retarded functional
  differential equations with applications, IEEE Transactions on Automatic
  Control 50~(8) (2005) 1090--1105.

\bibitem{impulsivo2007-3}
X.~Liu, Q.~Wang, The method of lyapunov functionals and exponential stability
  of impulsive systems with time delay, Nonlinear Analysis 66 (2007)
  1465--1454.

\bibitem{S2009}
I.~M. Stamova, Stability Analysis of Impulsive Functional Differential
  Equations, Walter de Gruyter, 2009.

\bibitem{Pri}
J.~A. Prieto, A.~Barreiro, S.~Dormido, S.~Tarbouriech, Delay-dependent
  stability of reset control systems with anticipative reset conditions, in:
  IFAC Symposium on Robust Control Design, 2012, pp. 219--224.

\bibitem{Guo}
Y.~Guo, L.~Xie, Quadratic stability of reset control systems with delays, in:
  10th World Congress on Intelligent Control and Automation (WCICA), 2012, pp.
  2268--2273.

\bibitem{DavBan}
M.~A. Dav{\'o}, A.~Ba{\~n}os, Delay-dependent stability of reset control
  systems with input/output delays, in: IEEE 52nd Annual Conference on Decision
  and Control (CDC), 2013, pp. 2018--2023.

\bibitem{MerDav}
P.~Mercader, M.~A. Dav{\'o}, A.~Ba{\~n}os, $\mathcal{H}_\infty$/$\mathcal{H}_2$
  analysis for time-delay reset control systems, in: 3rd International
  Conference on Systems and Control (ICSC), 2013, pp. 518--523.

\bibitem{MerCar}
P.~Mercader, J.~Carrasco, A.~Ba{\~n}os, {IQC} analysis for time-delay reset
  control systems with first order reset elements, in: IEEE 52nd Annual
  Conference on Decision and Control (CDC), 2013, pp. 2251--2256.

\bibitem{CarBanArj}
J.~Carrasco, A.~Ba{\~n}os, A.~van~der Schaft, A passivity-based approach to
  reset control systems stability, Systems \& Control Letters 59~(1) (2010)
  18--24.

\bibitem{teel}
J.~Liu, A.~Teel, Generalized solutions to hybrid systems with delays, in:
  Proceedings of the 51st Annual Conference on Decision and Control, 2012, pp.
  6169--6174.

\bibitem{teel2}
J.~Liu, A.~Teel, Hybrid systems with memory: Modelling and stability analysis
  via generalized solutions, in: Proceedings of the 19th {IFAC} World Congress,
  2014.

\bibitem{r_BPTZ2014}
A.~Ba{\~n}os, F.~Perez, S.~Tarbouriech, L.~Zaccarian, Low-Complexity
  Controllers for Time-Delay Systems: Delay-independent stability via reset
  loops, Advances in Delays and Dynamics, Springer International Publishing,
  2014, Ch.~8, pp. 111--125.

\bibitem{c13}
R.~Goebel, A.~R. Teel, R.~G. Sanfelice, Hybrid Dynamical Systems: Modeling,
  Stability, and Robustness, Princeton University Press, 2012.

\end{thebibliography}

\end{document}